# Moving MRI: Imaging a moving body with a moving magnet

## Authors

Jingting Yao,[1,2]* Artan Kaso,[1,2,3] Nikhil Patel,[1] Yin-Ching Iris Chen,[1,2] Andre van der Kouwe,[1,2] Daniel M. Merfeld,[4] Jerome L. Ackerman[1,2]

## Affiliations

[1]Athinoula A. Martinos Center for Biomedical Imaging, Massachusetts General Hospital, Charlestown, MA, USA, 02129.

[2]Department of Radiology, Harvard Medical School, Boston, MA, USA, 02115.

[3]Present address: SCAN Center, Stony Brook University, Stony Brook, NY, USA 11794.

[4]Ohio State University College of Medicine, Columbus, OH, USA 43210.

*Corresponding author (jy1015@mgh.harvard.edu)

## Abstract

Current magnetic resonance imaging (MRI) requires the subject to remain stationary to limit motion artifacts and avoid unwanted field-induced brain stimulation. However, imaging during large-scale motion could enable studies in which motion itself is central. One example is the study of brain networks involved in vestibular function, which senses head motion. Here, we demonstrate Moving MRI (mMRI), a system that enables imaging during large-scale motion by moving the subject and scanner together to minimize relative motion. We implemented a proof-of-concept platform using a compact, cryogen-free superconducting magnet mounted on a pneumatically actuated tilt mechanism that moves the magnet, gradients, and RF coil as a unit during scanning. Phantom and *in vivo* rat brain scans were acquired during repetitive tilting. We characterized artifacts arising from tilt-induced field shifts and residual subject–scanner motion, and partially reduced these effects. mMRI enables imaging during large-scale movement and may broaden access to naturalistic vestibular paradigms while providing a foundation for future human systems.

## Teaser

MRI during motion becomes possible by moving the scanner together with the subject.

## Introduction

Magnetic Resonance Imaging (MRI), including functional MRI (fMRI), is a powerful noninvasive tool used extensively in mapping brain anatomy and function[1,2]. Conventional MRI requires both the subject's body and the magnet, including the radio-frequency (RF) and gradient coils, to remain stationary to minimize motion-related image degradation and avoid field-induced physiological effects on the subject[3-5]. For these reasons, standard neuroimaging generally restricts subjects to minimal motion during scanning. In fMRI studies involving visual, tactile, or auditory stimulation, the need to minimize subject motion necessarily restricts experimental paradigms to relatively static conditions[6]. This limitation hinders investigation of neural processing during naturalistic, large-amplitude head and/or body motion, in which vestibular, autonomic, and other sensorimotor signals are intrinsically coupled to motion[7-11].

To address this limitation, we introduce moving MRI (mMRI), in which the magnet and subject’s head are spatially coupled and move in synchrony (**Fig. 1a,b**). Rather than moving the subject within the bore of a stationary magnet, mMRI applies the motion paradigm to the entire imaging assembly—magnet, gradient and shim coils, RF coil, and subject—thereby reducing relative head–scanner motion as a major source of artifacts. This approach has become feasible because recent superconducting magnet technology has enabled MRI-quality cryogen-free magnets, in which the windings are conduction-cooled by an electrically powered cryocooler, eliminating the need for liquid helium or nitrogen. These systems can be safely tilted and moved while at field, while maintaining sufficient field stability for MR measurements. Here, we demonstrate a proof-of-concept mMRI implementation using a compact 1.5-Tesla (T) cryogen-free extremity (arm and leg) magnet (**Fig. 1c** and **Methods: Extremity MRI Scanner**) on a pneumatically actuated tilt mechanism (**Fig. 1d–f**), with computer-controlled motion (**Methods: Automatic Motion Control**).

Despite the growing development of portable[12-15] and head-mounted[16,17] MRI systems where the magnet and head remain stationary during imaging, the potential for imaging during magnet motion has remained largely untapped. At present, methods established for human studies during large-scale head motion include magnetoencephalography with optically pumped magnetometers[18] and electroencephalography[19], both of which offer relatively low spatial resolution, and functional near-infrared spectroscopy[20], which is limited to imaging the cortical surface. mMRI therefore holds the promise of introducing high-quality, information-rich imaging to the study of brain activation during motion.

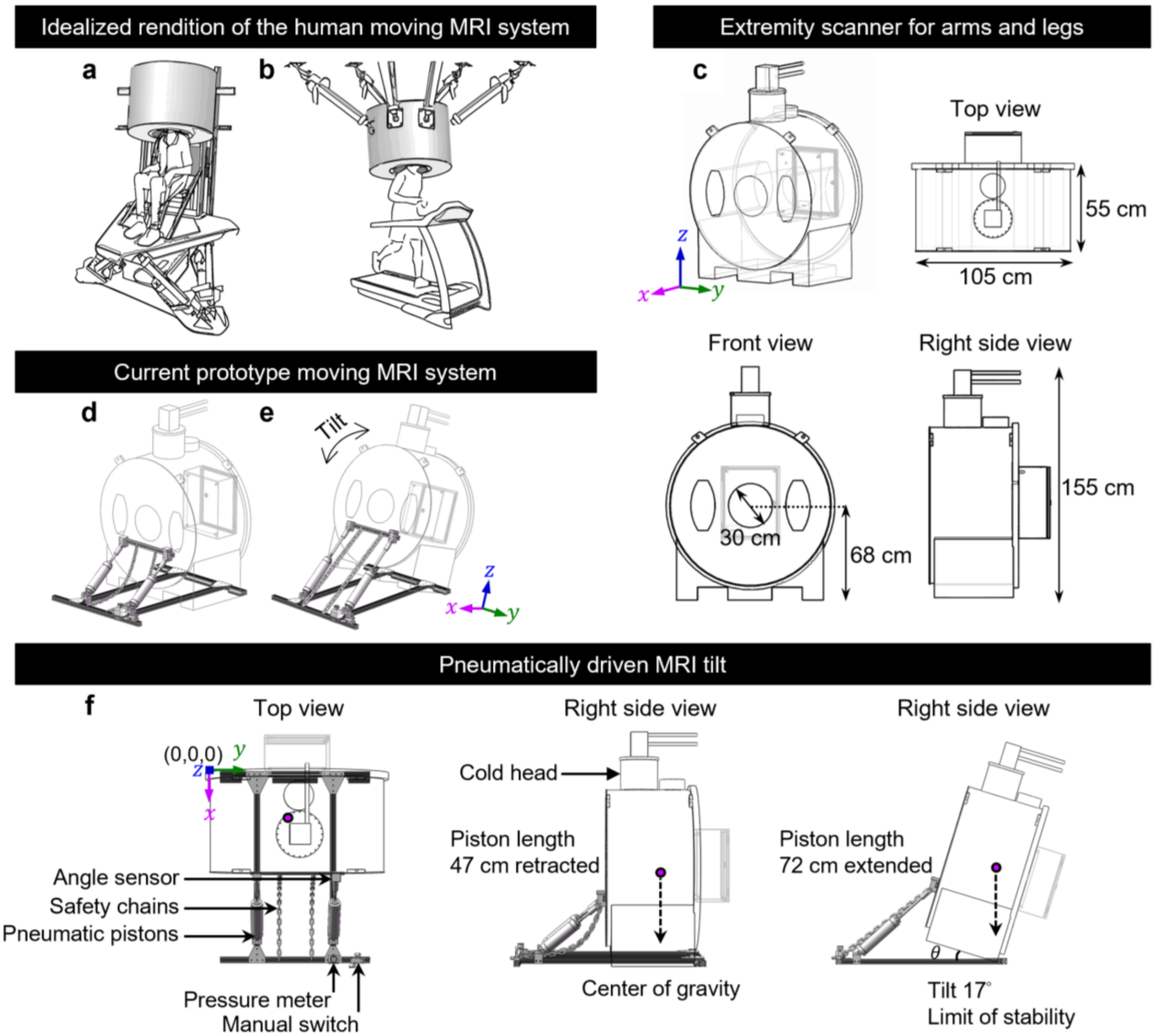

**Fig. 1 Conceptual design and current implementation of Moving MRI (mMRI). a** Idealized human-scale mMRI concept in which the subject and the magnet are coupled and move synchronously, minimizing relative motion between the head and scanner during imaging. **b** Alternative human implementation with the subject seated on a motion platform capable of controlled motion in multiple degrees of freedom, potentially using compliant robotic actuation to provide position and force/torque feedback. **c** Compact extremity MRI scanner geometry used as the basis for the present study; top, front, and right-side views are shown with key dimensions (overall footprint 105 × 55 cm; height 155 cm; central bore diameter 30 cm). **d, e** Current prototype mMRI scanner shown in the baseline and tilted configurations during repetitive tilt motion; the coordinate axes indicate the reference frame used in this work. **f** Pneumatically actuated tilt mechanism: top view showing the pneumatic pistons, angle sensor,

safety chains, and manual pressure control; right-side views show the retracted (47 cm) and extended (72 cm) piston lengths, the resulting tilt angle (17°), and the center-of-gravity location relative to the stability limit. Note that for the purpose of computing the field shifts with tilting, the z-axis is defined as Earth-vertical when the magnet is horizontal, rather than the direction of $B_0$.

mMRI is expected to be especially relevant for assessing vestibular dysfunction[21-24], because vestibular sensation naturally responds to head motion. In natural behavior, canal- and otolith-driven signals are tightly coupled to real head motion and integrated with visual and proprioceptive cues. In contrast, existing vestibular MR imaging studies rely on artificial stimuli, such as heat, cold, sound and electric current, that introduce confounding co-stimulation sensations and do no engage the brain in the same way as natural motion[25,26]. The study of vestibular function with modern MR imaging methods remains challenging due to the need for the imaged subject to remain nearly stationary relative to the magnet during scanning. Nonetheless, recent advancements in vestibular stimulation, including galvanic vestibular stimulation[27], caloric vestibular stimulation[28], hexapod motion platforms[21], and rotatory chairs[29,30], carry the growing potential of combining motion-based vestibular paradigms with imaging methods such as fMRI. Beyond vestibular applications, mMRI may also open up new avenues for research in various other areas, including traumatic brain injury[31,32], human motor control[33], and spinal compression[34].

In this study, we developed a prototype mMRI system and demonstrated phantom and *in vivo* rat brain imaging during repetitive magnet tilting. Image degradation was dominated by tilt-dependent field shifts and residual subject–scanner motion, and we applied postprocessing corrections to mitigate these artifacts. The field shifts were driven primarily by weak background magnetic fields in the laboratory, including the Earth's geomagnetic field and fringe fields from a nearby 14 T vertical MR magnet. Fortuitously, the 14 T magnet was decommissioned during the course of the study, creating a natural experiment in which we could compare data acquired with the 14 T fringe field present versus absent. This allowed us to confirm the fringe field as a primary source of the observed tilt-dependent field shifts.

## Results

Results are reported for two conditions: (1) with a nearby actively shielded 14 T (proton frequency 600 MHz) MRI magnet energized ("14 T ON") and (2) with the 14 T magnet de-energized ("14 T OFF"). The site layout is presented in (**Methods: Extremity MRI Scanner**). All measurements were performed over tilt angles up to 17°, the maximum mechanically stable tilt of the extremity magnet as determined by a center-of-gravity analysis (**Methods: Center-of-Gravity and Maximum Tilt Angle** and **Supplementary Materials A**).

### *Tilt-Angle Dependence of Resonance Characteristics*

Resonance behavior was characterized as a function of scanner tilt angle under four measurement conditions, with the 14 T magnet either ON or OFF and with measurements acquired under both static and dynamic tilting conditions. For each resonance spectrum, the resonance frequency shift, Δf, was estimated from the peak location, and the linewidth and normalized peak magnitude were extracted. A frequency shift of 0 Hz corresponded to a resonant frequency of 63.87 MHz when the magnet was in the baseline untilted position. During static conditions, the scanner was held at a fixed tilt angle for several minutes such that measurements were not affected by mechanical, electrical or magnetic transients.

Under the 14 T ON condition, a pronounced and monotonic shift of the resonance peak toward positive Δf was observed with increasing tilt angle in static measurements (**Fig. 2a,c**). The Δf–angle relationship was found to be highly repeatable across runs and was well described by a linear model over the tested range, with slopes on the order of ~18 Hz/deg and coefficients of determination exceeding 0.99 (**Fig. 2c**). Over an angle excursion from 0° to 17°, a maximum shift of 312 Hz occurred, corresponding to 4.89 ppm at 63.87 MHz. A gradual increase in linewidth with tilt angle was observed (**Fig. 2d**), consistent with broadened resonance associated with degraded shimming. A systematic reduction in peak magnitude with increasing angle was also observed, with greater run-to-run variability than was seen for Δf and linewidth (**Fig. 2e**).

During dynamic measurements under 14 T ON condition, similar trends were observed but with broader spectra and a steeper Δf–angle dependence (**Fig. 2b,f–h**). The resonance peak shifted to higher positive Δf at larger angles than in the static condition (**Fig. 2b**), and fitted slopes increased to approximately 25–26 Hz/deg (**Fig. 2f**). Linewidths measured during motion increased with angle but were elevated relative to static measurements and (**Fig. 2g**), consistent with additional broadening under time-varying conditions. Peak magnitudes during dynamic measurements were largely preserved across angles (**Fig. 2h**). The dynamic Δf–angle dependence was further evaluated by repeatedly acquiring free-induction decays while the scanner was tilted back and forth.

In the 14 T OFF condition, the tilt dependence differed. In static measurements, the resonance peak shifted toward negative Δf with increasing tilt angle (**Fig. 2a,i**), and the full-range dependence was captured more accurately by a weakly nonlinear (quadratic) trend (**Fig. 2i**). In contrast to 14 T ON condition, linewidths remained approximately constant across angles (**Fig.**

**2j**), and peak magnitudes decreased only modestly and reproducibly with angle (**Fig. 2k**). During dynamic measurements under the 14 T OFF condition, increased scatter in Δf was observed relative to the static case (**Fig. 2l**), whereas linewidths remained within a narrow range (**Fig. 2m**) and peak magnitudes were largely maintained (**Fig. 2n**).

RF coil performance was monitored during magnet motion. Under both 14 T ON and OFF conditions, the return loss was −40.03 ± 3.44 dB (8.6% relative variability) and the loaded quality factor was 48.18 ± 0.50 (1.0% relative variability) across the tested angles.

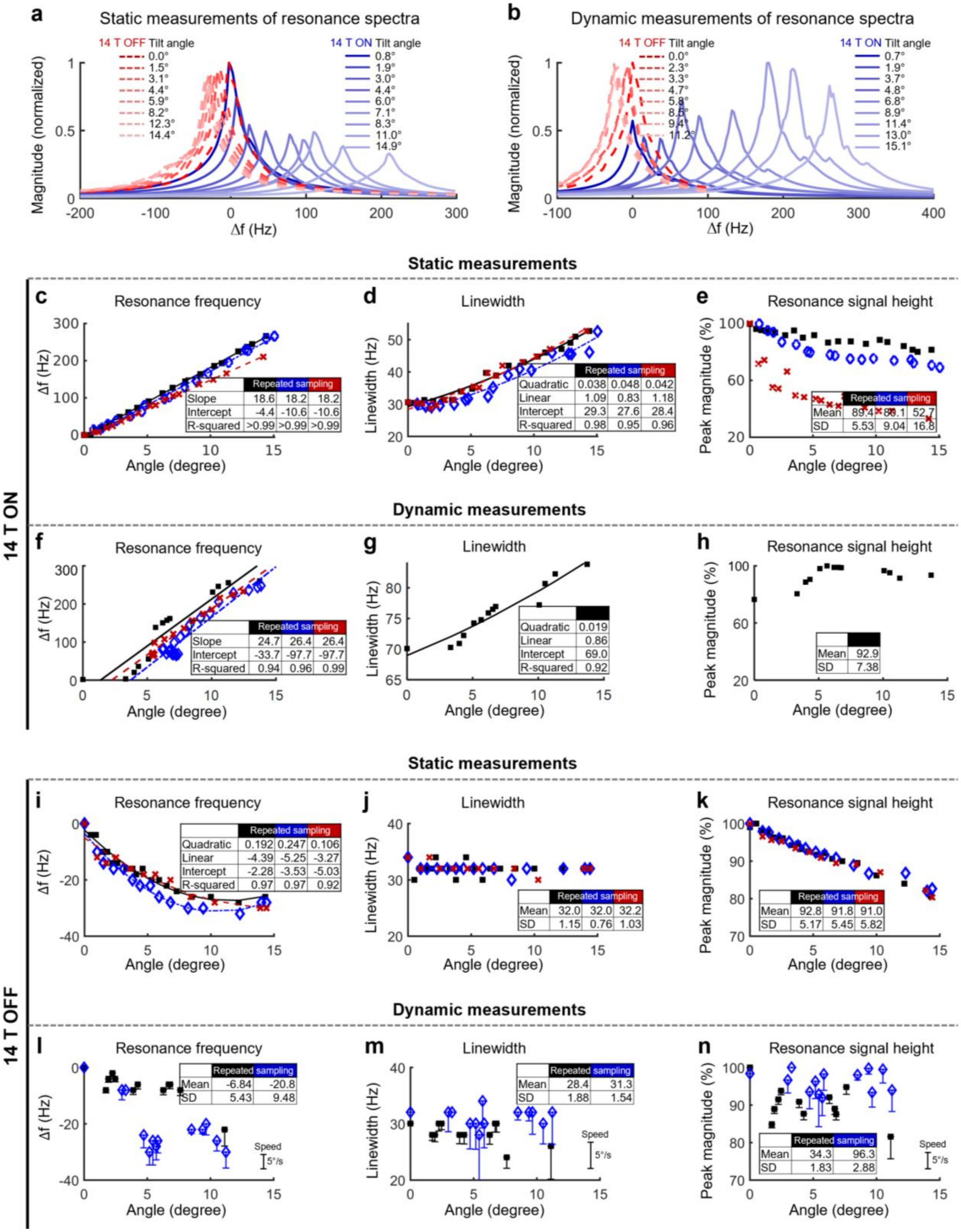

a
Static measurements of resonance spectra
14 T OFF Tilt angle
0.0°
1.5°
3.1°
4.4°
5.9°
8.2°
12.3°
14.4°
14 T ON Tilt angle
0.8°
1.9°
3.0°
4.4°
6.0°
7.1°
8.3°
11.0°
14.9°
Magnitude (normalized)
Δf (Hz)
b
Dynamic measurements of resonance spectra
14 T OFF Tilt angle
0.0°
2.3°
3.3°
4.7°
5.8°
8.5°
9.4°
11.2°
14 T ON Tilt angle
0.7°
1.9°
3.7°
4.8°
6.8°
8.9°
11.4°
13.0°
15.1°
Magnitude (normalized)
Δf (Hz)
14 T ON
Static measurements
c
Resonance frequency
Δf (Hz)
Angle (degree)
Repeated sampling
Slope 18.6 18.2 18.2
Intercept -4.4 -10.6 -10.6
R-squared >0.99 >0.99 >0.99
d
Linewidth
Linewidth (Hz)
Angle (degree)
Repeated sampling
Quadratic 0.038 0.048 0.042
Linear 1.09 0.83 1.18
Intercept 29.3 27.6 28.4
R-squared 0.98 0.95 0.96
e
Resonance signal height
Peak magnitude (%)
Angle (degree)
Repeated sampling
Mean 89.4 80.1 52.7
SD 5.53 9.04 16.8
Dynamic measurements
f
Resonance frequency
Δf (Hz)
Angle (degree)
Repeated sampling
Slope 24.7 26.4 26.4
Intercept -33.7 -97.7 -97.7
R-squared 0.94 0.96 0.99
g
Linewidth
Linewidth (Hz)
Angle (degree)
Quadratic 0.019
Linear 0.86
Intercept 69.0
R-squared 0.92
h
Resonance signal height
Peak magnitude (%)
Angle (degree)
Mean 92.9
SD 7.38
14 T OFF
Static measurements
i
Resonance frequency
Δf (Hz)
Angle (degree)
Repeated sampling
Quadratic 0.192 0.247 0.106
Linear -4.39 -5.25 -3.27
Intercept -2.28 -3.53 -5.03
R-squared 0.97 0.97 0.92
j
Linewidth
Linewidth (Hz)
Angle (degree)
Repeated sampling
Mean 32.0 32.0 32.2
SD 1.15 0.76 1.03
k
Resonance signal height
Peak magnitude (%)
Angle (degree)
Repeated sampling
Mean 92.8 91.8 91.0
SD 5.17 5.45 5.82
Dynamic measurements
l
Resonance frequency
Δf (Hz)
Angle (degree)
Repeated sampling
Mean -6.84 -20.8
SD 5.43 9.48
Speed
5°/s
m
Linewidth
Linewidth (Hz)
Angle (degree)
Repeated sampling
Mean 28.4 31.3
SD 1.88 1.54
Speed
5°/s
n
Resonance signal height
Peak magnitude (%)
Angle (degree)
Repeated sampling
Mean 34.3 96.3
SD 1.83 2.88
Speed
5°/s

**Fig. 2 Resonance signal characteristics during magnet tilting. a** Static measurements of normalized resonance spectra acquired with the system held at discrete tilt angles. Spectra are shown for the two measurement conditions: 14 T ON (blue) and 14 T OFF (red). **b** Dynamic measurements of the normalized resonance spectra acquired during continuous tilting; spectra are shown by the instantaneous tilt angle at which they were sampled. **c–e** Resonance characterization from static measurements under 14 T ON condition: **c** resonance frequency shift (Δf; peak location) versus tilt angle, **d** linewidth, and **e** normalized peak magnitude. **f–h** Resonance characterization from dynamic measurements under the 14 T ON condition: **f** Δf versus tilt angle, **g** linewidth, and **h** peak magnitude. **i–k** Resonance characterization from static measurements under the 14 T OFF condition: **i** Δf versus tilt angle, **j** linewidth, and **k** peak magnitude. **l–n** Resonance characterization from dynamic measurements under the 14 T OFF condition: **l** Δf versus tilt angle, **m** linewidth, and **n** peak magnitude**.** Markers denote repeated measurements acquired on separate acquisitions (color/marker key in each panel). Solid curves indicate least-squares fits (linear or quadratic, as shown in insets), with fit parameters and coefficients of determination reported in-panel.

The tilt-dependent frequency shifts arise from reorientation of the scanner relative to static magnetic fields in the laboratory. These fields include the Earth's geomagnetic field as modified by building materials and nearby magnetic objects, as well as, when present, the fringe field of the nearby 14 T magnet. Tilting changes the projection of this combined background field onto the 1.5 T main field, producing a corresponding shift in resonance frequency. The relative contributions differ between the 14 T ON and OFF conditions, with the fringe field dominating when present and the ambient laboratory field governing the response otherwise. These magnitudes are consistent with expectations for projection of external fields on the order of tens of microtesla (Earth field) to sub-millitesla (fringe field) onto the scanner's main field direction.

***Magnet Tilt–Induced Image Artifacts***

To characterize how tilt-induced frequency shifts manifest as image artifacts, a phantom was scanned while the 1.5 T magnet was at rest in the horizontal position, and during repeated up/down tilting, under both 14 T ON and 14 T OFF conditions (**Methods: Phantom mMRI Experiments**, **Fig. 3**). During acquisition over multiple tilt cycles (up to 5.9°/s cycling at 0.02 Hz), successive phase-encoding lines were acquired at slightly different tilt angles, causing inconsistencies in frequency reference between lines and resulting image artifacts. Retrospective correction of individual phase-encoding lines based on projection displacement along the frequency encoding direction substantially improved image quality.

To characterize the time dependence of the resonance frequency changes during scanning, an “image” was acquired with the phase-encoding gradient disabled, yielding a time series of one-dimensional projections of the phantom along the frequency-encoding axis[35]. The recorded tilt-angle time course was smoothed with a Gaussian moving filter (window size 50) to suppress high-frequency field variations from the magnet cold head (~4 Hz resonance frequency shift occurring at a ~2 Hz cycle). Under these conditions, the projections exhibited shifts along the frequency-encoding axis that strongly correlated with tilt angle (Pearson’s $r = 0.94$, $p < 0.0001$) for the 14 T ON condition, following a linear relationship consistent with that shown in **Fig. 2f**.

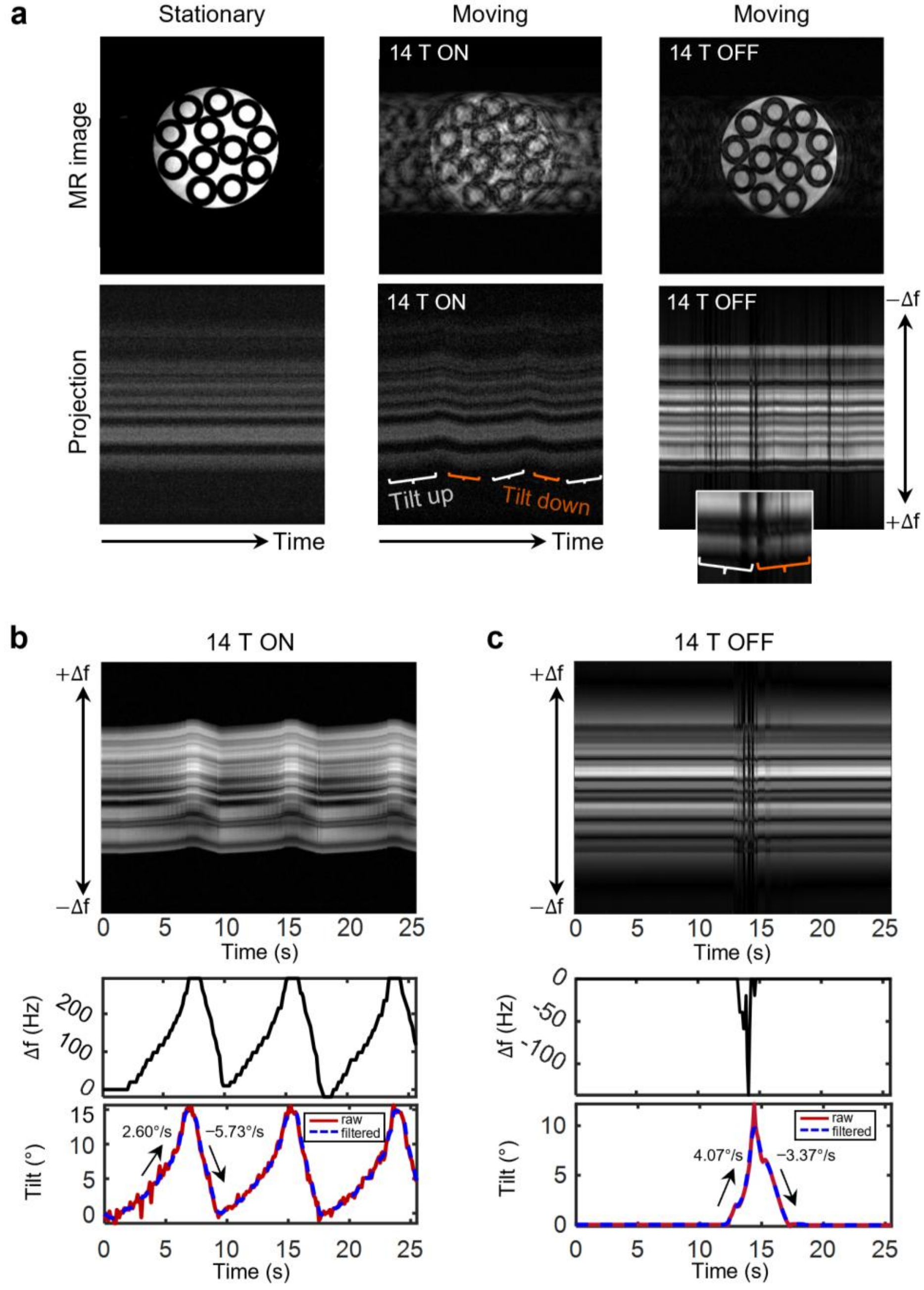

**Fig. 3 Tilt-induced frequency shifts and image artifacts in phantom MRI. a** Phantom MR images (top row) and corresponding one-dimensional projections along the frequency-encoding axis (bottom row), displayed over time, acquired with the magnet stationary (left) and during dynamic tilting with the 14 T magnet ON (middle) and OFF (right). During motion, the projections shift with tilt, producing apparent displacement along the frequency-encoding axis. With the 14 T OFF, this displacement is reduced (inset). **b** With the 14 T ON, sequential projections over time (top) show repeated banded displacement across tilt cycles. The corresponding resonance-frequency shift Δf(t) (middle) and tilt angle (bottom; raw and Gaussian-filtered traces) exhibit a strong, nearly linear relationship. **c** With the 14 T OFF, a representative single tilt cycle shows markedly smaller frequency excursions in both the stacked projections (top) and $\Delta f(t)$ trace (middle), consistent with the reduced tilt–frequency dependence in the absence of the 14 T fringe field.

***Tilt-Informed Artifact Correction***

Nominal retrospective correction (**Methods: Correction for Tilt-Induced Frequency Shifts; Supplementary Materials B**) was applied using the per-line frequency shift, $\Delta f(k_y)$, estimated from the recorded tilt angle associated with each acquired frequency-encoding line $k_y$, together with the static angle–frequency relationship $\Delta f(\theta)$ reported in **Fig. 2c** (14 T ON) and **Fig. 2i** (14 T OFF)**.** Correction performance was evaluated using predefined regions of interest (ROIs) for signal, noise, and artifact (**Fig. 4c–d**), and quantified using metrics computed from the reconstructed magnitude image, $|I|$, where $I$ denotes the reconstructed complex image. Artifact-SD ($\sigma_{art}$) was defined as the standard deviation of $|I|$ within the artifact ROI. Signal-Mean ($\mu_{sig}$) and Artifact-Mean ($\mu_{art}$) were defined as the mean of $|I|$ within the signal and artifact ROIs, respectively. Noise-SD ($\sigma_n$) was defined within the noise ROI. The signal-to-noise ratio (SNR) was calculated as $\mu_{sig}/\sigma_n$, and the artifact-to-noise ratio (ANR) as $\mu_{art}/\sigma_n$.

**Figure 4** and **Table 1** illustrate nominal tilt-informed correction under the 14 T ON condition for a phantom scan (using a spin-echo pulse sequence) and a live rat scan (using a gradient-echo pulse sequence). In the phantom (**Fig. 4a**), motion produced prominent phase-encoding ghosting and intensity modulation, with SNR decreasing from 68.0 (stationary) to 53.2 (moving, uncorrected) and ANR increasing from 2.20 to 14.9. **Table 1** shows that these changes occurred with a nearly unchanged noise background (Noise-SD: 0.65–0.66), alongside reduced signal intensity (Signal-Mean: 44.2 to 35.1) and increased artifact variability (Artifact-SD: 0.86 to 1.13). After nominal correction, SNR recovered to 64.8 (≈95% of the stationary value), and ANR decreased to 10.3; Signal-Mean improved to 42.8 and Artifact-SD decreased to 0.99. The constancy of the noise background demonstrates the stability of the scanner electronics with tilting, and that the observed changes reflect the motion-induced artifacts.

In the rat experiment (**Fig. 4b**), motion reduced SNR from 33.4 to 28.4 and increased ANR from 2.07 to 4.17. Noise-SD again remained stable (0.10–0.11) (**Table 1**), while Artifact-SD increased from 0.11 to 0.34, with only a modest decrease in Signal-Mean (3.29 to 3.02). After correction, ANR decreased to 3.38 and SNR increased to 30.8, with reduced Artifact-SD (0.22) and unchanged Noise-SD.

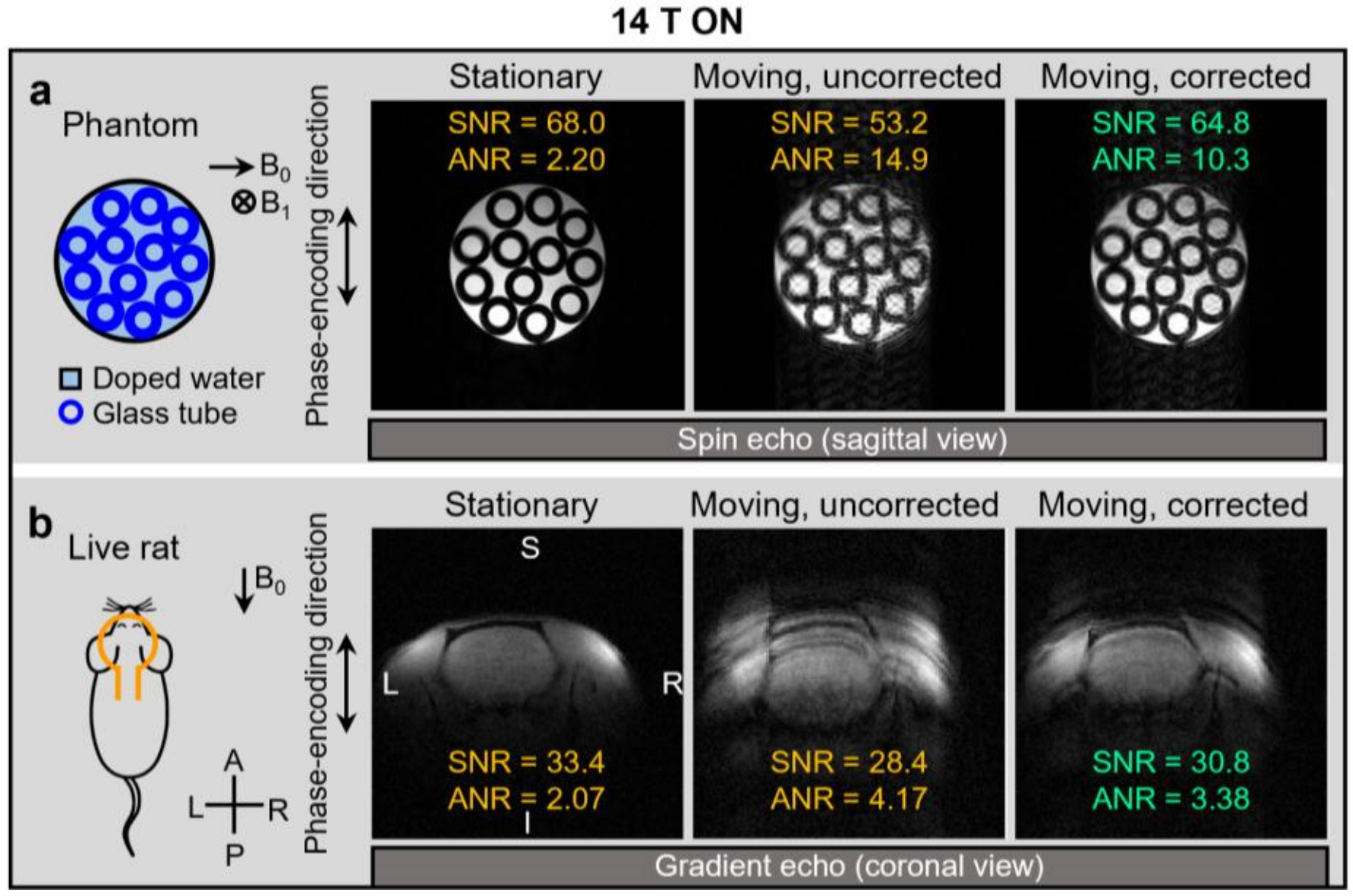

**Fig. 4 Nominal retrospective correction for tilt-induced frequency offsets (14 T ON). a** Structured phantom (doped water with glass tubes), spin-echo sagittal images shown for stationary, moving (uncorrected), and moving after nominal correction. Motion produces phase-encoding ghosting and structured artifacts; nominal correction partially restores contrast and suppresses artifacts. SNR and ANR values are overlaid on each image. **b** Live rat, gradient-echo coronal images shown for stationary, moving (uncorrected), and moving after nominal correction. **c–d** Predefined regions of interest used for quantitative metrics: Signal ROI (green), Artifact ROI (yellow) in a signal-free region, and Noise ROI (white rectangle). A fixed intensity-based mask was applied to define the phantom signal ROI, excluding the low-signal glass-tube cores.

**Table 1**. ROI-based image-quality metrics for phantom and rat experiments (14 T ON)

| Object | Metric | Static | Moving | Corrected |
|---|---|---|---|---|
| **Phantom** | $\mu_{sig} \pm \sigma_{sig}$ | 44.2 ± 1.54 | 35.1 ± 1.42 | 42.8 ± 1.32 |
| | $\sigma_n$ | 0.65 | 0.66 | 0.66 |
| | $\sigma_{art}$ | 0.86 | 1.13 | 0.99 |
| **Rat** | $\mu_{sig} \pm \sigma_{sig}$ | 3.29 ± 0.76 | 3.02 ± 0.72 | 3.13 ± 0.66 |
| | $\sigma_n$ | 0.10 | 0.11 | 0.11 |
| | $\sigma_{art}$ | 0.11 | 0.34 | 0.22 |

μ and σ indicate mean and standard deviation.

***Mask-Optimized Artifact Correction***

Residual artifact persisted after nominal correction in some acquisitions, motivating a constrained refinement of $\Delta f(k_y)$ guided by a predefined artifact mask (**Methods: Mask-Optimized Artifact Correction**). With 14 T ON (**Fig. 5**), nominal correction reduced visible residual ghosting but yielded limited or inconsistent decreases in $\sigma_{art}$ (e.g., slow: $\sigma_{art}$ = 0.121 vs. 0.118 uncorrected; fast: 0.151 vs. 0.199 uncorrected). Mask-optimized refinement, using center-out, top-down, or random update orders, consistently reduced $\sigma_{art}$ toward the stationary reference (stationary: $\sigma_{art}$ = 0.107; fast: 0.107–0.110 after refinement). A similar trend was observed with 14 T OFF (**Fig. 6**), where nominal correction failed at higher tilt rates (fast: $\sigma_{art}$ = 0.259, exceeding the uncorrected value of 0.242), whereas refinement strategies robustly reduced $\sigma_{art}$ (fast: 0.137–0.141).

Center-out ordering yielded the most stable reductions in $\sigma_{art}$. Updates near central $k_y$ had the largest effect on artifact suppression, pointing to the dominant role of low spatial frequencies in shaping overall image structure. Order dependence diminished when multiple sweeps were performed, as earlier updates were revisited and refined. With bounded offsets and limited sweeps (combined with random restarts), mask-optimized refinement further reduced $\sigma_{art}$ beyond nominal correction while maintaining $\mu_{sig}$ across correction variants, without evidence of global signal attenuation. Per-line optimization traces showed that $\sigma_{art}$ was most sensitive to adjustments near central $k_y$; this sensitivity explains the advantage of center-out updates.

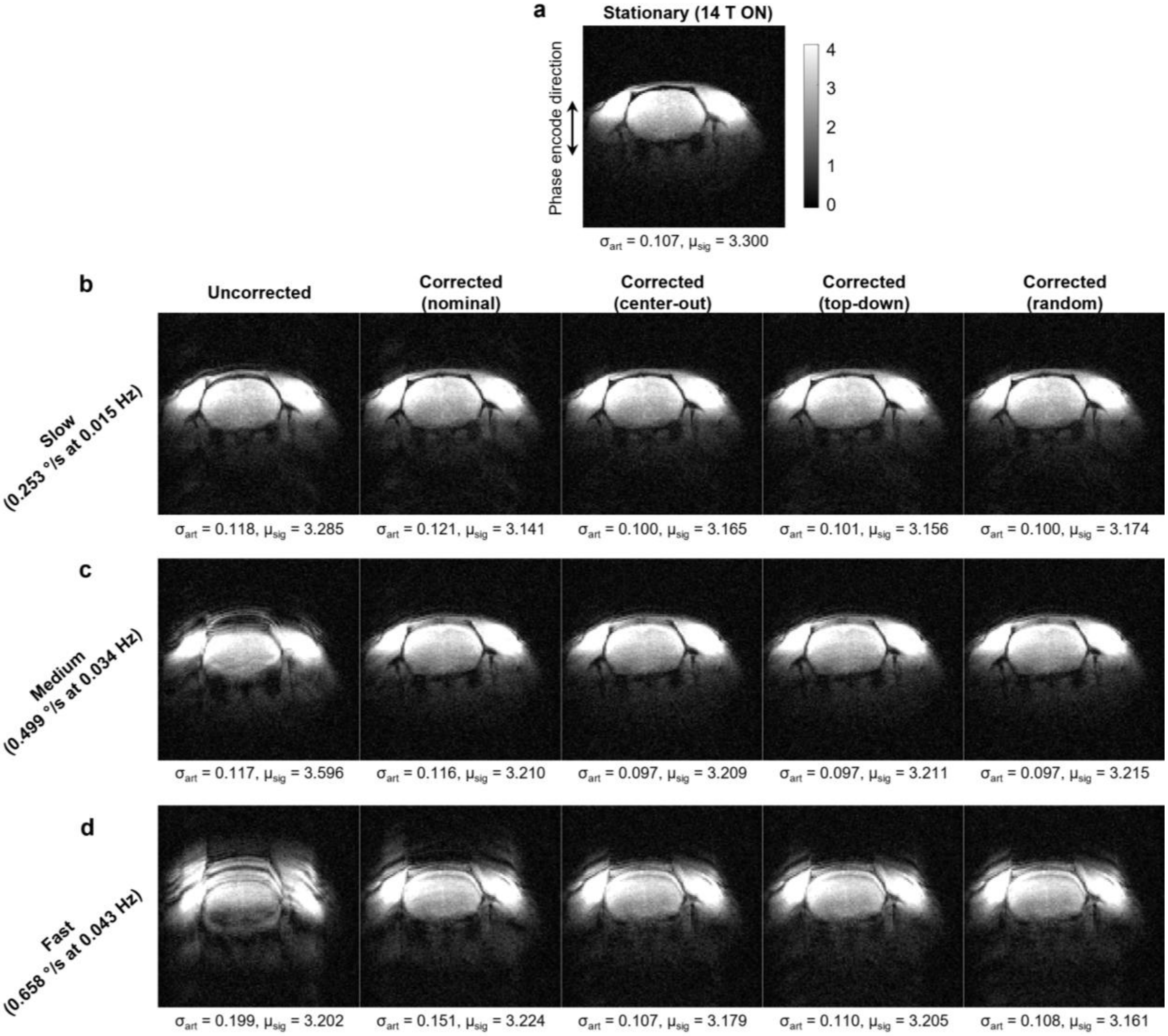

**Fig. 5 Mask-optimized artifact correction with 14 T ON. a** Stationary reference acquisition. $\sigma_{art}$ and $\mu_{sig}$ are reported for the artifact ROI and signal ROI, respectively. **b–d** Acquisitions at slow, medium, and fast tilt rates, shown as uncorrected, nominally corrected, and mask-optimized refinements using three $k_y$ update orders (center-out, top-down, random). Mask-optimized refinement reduced $\sigma_{art}$ relative to nominal correction while preserving $\mu_{sig}$ across tilt conditions.

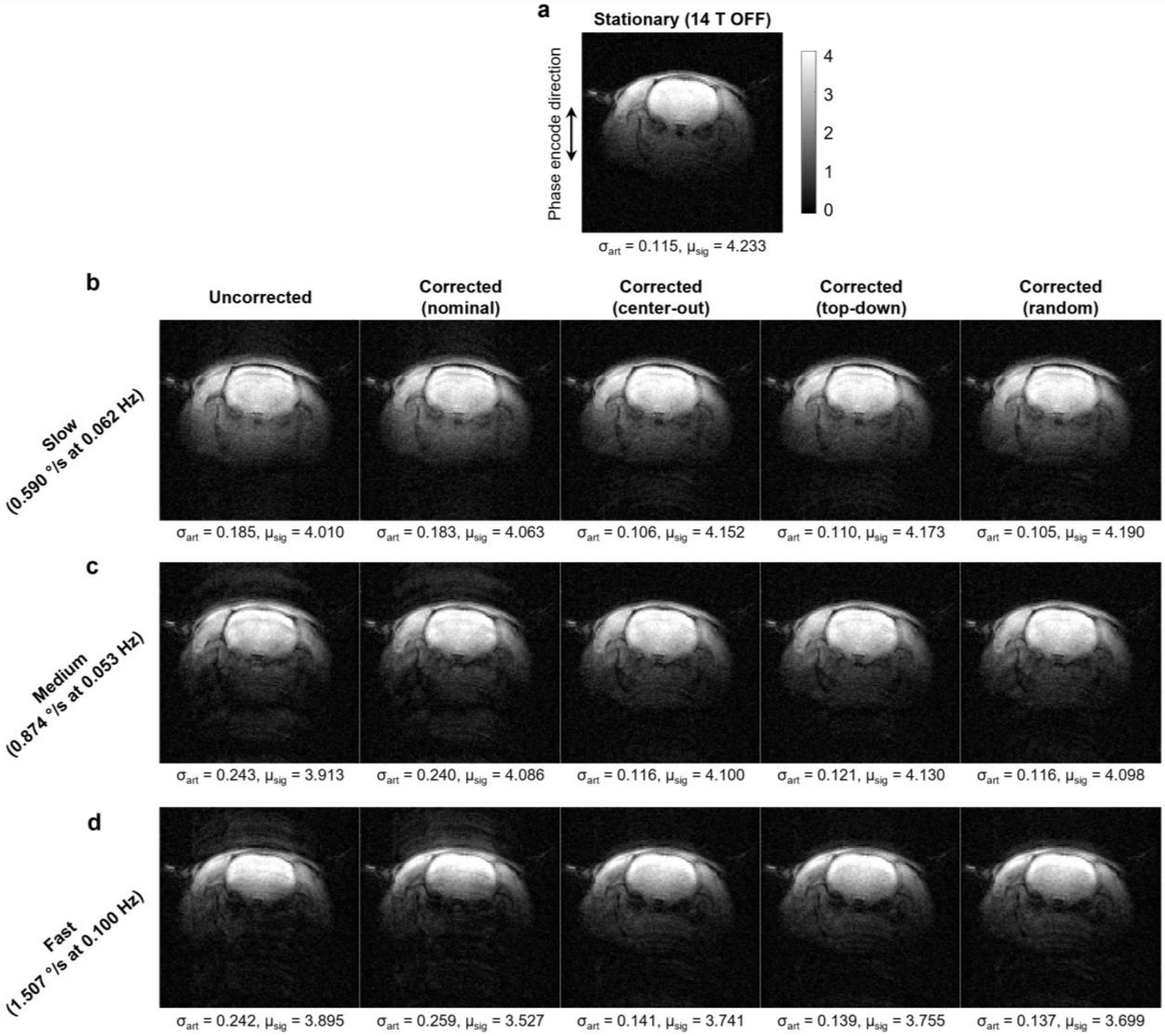

**Fig. 6 Mask-optimized artifact correction with 14 T OFF. a** Stationary reference acquisition. $\sigma_{art}$ and $\mu_{sig}$ are reported for the artifact ROI and signal ROI, respectively. **b–d** Acquisitions at slow, medium, and fast tilt rates, shown as uncorrected, nominally corrected, and mask-optimized refinements using three $k_y$ update orders (center-out, top-down, random). Mask-optimized refinement reduces $\sigma_{art}$ relative to nominal correction while preserving $\mu_{sig}$, and remains effective at higher tilt rates where nominal correction fails to reduce, and can increase, $\sigma_{art}$.

**Figures 7a–b** summarize artifact and signal metrics across conditions and correction strategies. Pair-wise testing using Wilcoxon signed-rank with Benjamini–Hochberg false discovery rate (FDR) correction showed significant lower Artifact-SD for all mask-optimized correction strategies (center-out, top-down, and random $k_y$ update orderings) relative to the uncorrected moving condition (all $p_{\mathrm{fdr}} = 0.01$), whereas the nominal correction did not make a statistically significant change (**Fig. 7a**). With 14 T OFF, Artifact-SD was similarly reduced across all correction strategies (**Fig. 7a**). Signal-Mean was comparable across stationary, uncorrected, and corrected datasets in both 14 T ON and OFF (**Fig. 7b**), suggesting that correction suppressed artifact without altering signal intensity.

Artifact-SD depended on both mean tilt velocity and the effective temporal sampling of the k-space (**Fig. 7c–d**). With 14 T ON, Artifact-SD grouped by repetition time (TR) generally increased with TR, particularly at higher tilt rates (**Fig. 7c**). However, the relationship was not solely determined by velocity. When TR was held constant (e.g., 14 T OFF, TR = 500 ms), Artifact-SD increased with tilt velocity (**Fig. 7d**).

Group-level statistics supported the dependence of Artifact-SD on $k_y$ update ordering strategy (**Table 2**). With 14 T ON (N = 8 datasets, averaged across 3 consecutive slices), Artifact-SD differed significantly among $k_y$ update orders (Kruskal–Wallis $p_{\mathrm{kw}} < 1\times10^{-8}$), and post-hoc testing with Benjamini–Hochberg FDR correction showed that nominal $k_y$ traversal resulted in higher Artifact-SD than center-out ($p_{\mathrm{fdr}} = 1.07\times10^{-6}$), top-down ($p_{\mathrm{fdr}} = 7.66\times10^{-6}$), and random ($p_{\mathrm{fdr}} = 4.19\times10^{-7}$) ordering. No differences were found among the three mask-optimized correction strategies (all $p_{\mathrm{fdr}} \approx 1$). With 14 T OFF (N = 5 single-slice datasets), the effect of $k_y$ update ordering remained significant (p = 0.01), but pairwise differences relative to nominal correction did not survive FDR correction (**Table 2**).

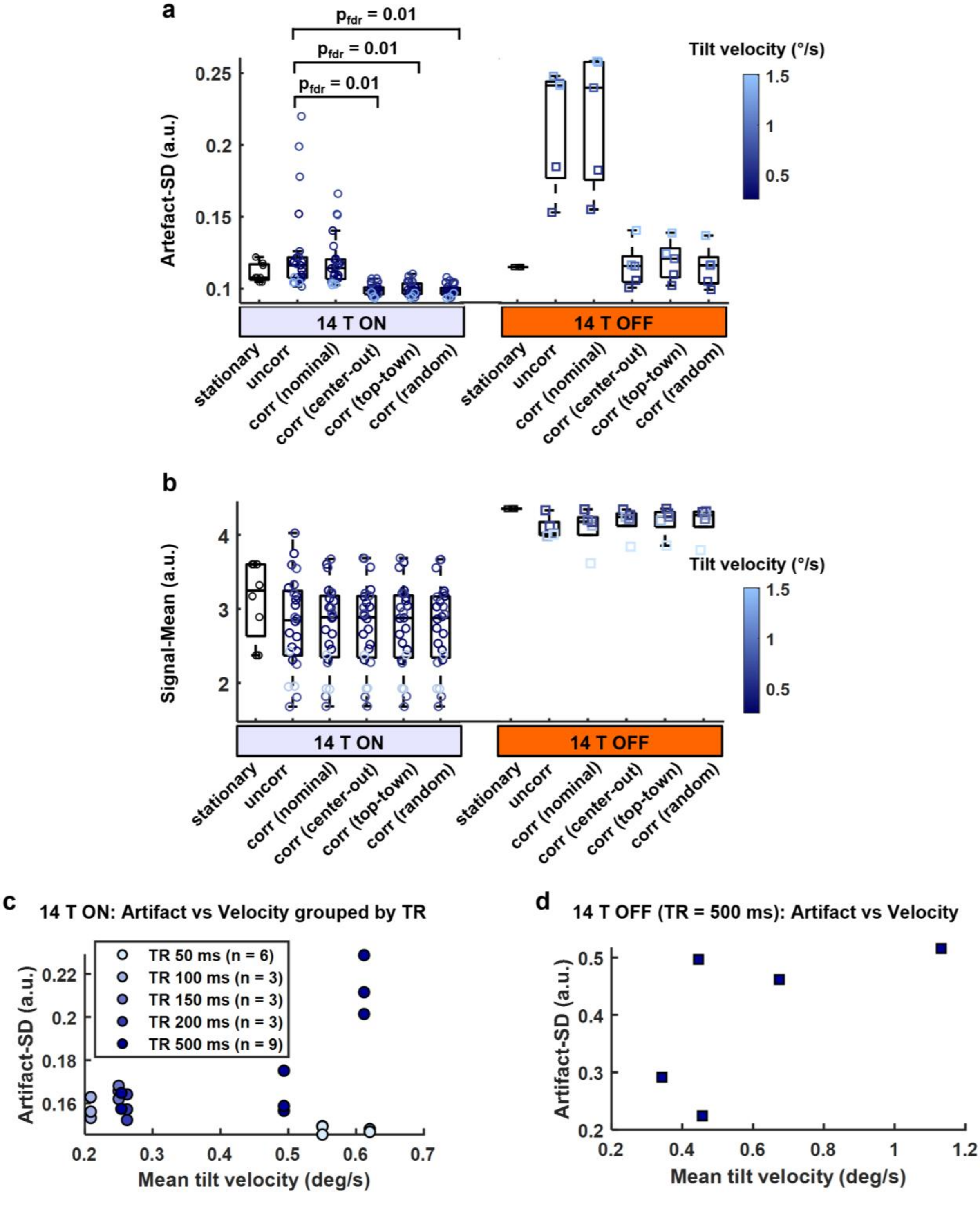

**Fig. 7 Summary of artifact and signal metrics across correction strategies and tilt velocity. a** Artifact-SD for stationary, uncorrected motion, nominal correction, and mask-optimized artifact correction (center-out, top-down, random $k_y$ update ordering) with 14 T ON and 14 T OFF. Each marker corresponds to a single acquisition, color-coded by mean tilt

velocity. Boxplots show the median (center line), interquartile range (25th–75th percentiles), and whiskers extending to 1.5× the interquartile range. Brackets indicate pairwise comparisons significant after Benjamini–Hochberg FDR correction. Markers show per-slice values; statistical testing was performed on dataset-level means (averaged across three slices per dataset for 14 T ON). **b** Signal-Mean for the same conditions shown in **a**. **c** 14 T ON: Artifact-SD versus mean tilt velocity, grouped by TR. In the legend, $n$ denotes the total number of slices in each TR group. **d** With 14 T OFF and TR = 500 ms, Artifact-SD generally increased with tilt velocity.

**Table 2 Artifact-SD comparison across artifact correction strategies**

| 14 T condition | Comparison | $p_{wsr}$ | $p_{\mathrm{fdr}}$ | $p_{kw}$ |
|---|---|---|---|---|
| **ON (N = 8)** | Nominal vs Center-out | $3.58\times10^{-7}$ | $1.07\times10^{-6}$ | $< 1\times10^{-8}$ |
| | Nominal vs Top-down | $3.83\times10^{-6}$ | $7.66\times10^{-6}$ | $< 1\times10^{-8}$ |
| | Nominal vs Random | $6.99\times10^{-8}$ | $4.19\times10^{-7}$ | $< 1\times10^{-8}$ |
| | Center-out vs Top down | 1.00 | 1.00 | $< 1\times10^{-8}$ |
| | Center-out vs Random | 1.00 | 1.00 | $< 1\times10^{-8}$ |
| | Top-down vs Random | 1.00 | 1.00 | $< 1\times10^{-8}$ |
| **OFF (N = 5)** | Nominal vs Center-out | 0.03 | 1.00 | 0.01 |
| | Nominal vs Top-down | 0.11 | 0.21 | 0.01 |
| | Nominal vs Random | 0.02 | 1.00 | 0.01 |
| | Center-out vs Top down | 1.00 | 1.00 | 0.01 |
| | Center-out vs Random | 1.00 | 1.00 | 0.01 |
| | Top-down vs Random | 1.00 | 1.00 | 0.01 |

N denotes the number of datasets. With 14 T ON, each dataset is the mean of 3 consecutive slices. With 14 T OFF, each dataset is from a single-slice acquisition. $p_{\mathrm{wsr}}$, $p_{\mathrm{fdr}}$ and $p_{kw}$ represent two-sided paired Wilcoxon signed-rank test, Benjamini–Hochberg false discovery rate correction, and Kruskal–Wallis test, respectively.

## Discussion

This work demonstrates the feasibility of acquiring MRI data during controlled, large-amplitude motion using a compact 1.5 T cryogen-free extremity magnet, enabled by physically moving the magnet together with the imaged object. This motion paradigm is fundamentally distinct from moving the subject alone along a comparable large-scale spatial trajectory within a stationary magnet. After correcting for the dominant tilt-dependent global frequency shift, image quality improved substantially, but the remaining artifacts also reveal limits of the $0^{th}$-order correction. These artifacts persist when the tilt-induced field perturbation varies spatially, contains higher-order frequency deviations, or evolves on timescales not adequately sampled by the acquisition (e.g., long TR and slow sampling), indicating that a single per-line correction based solely on the static frequency–tilt relationship is insufficient.

mMRI introduces sources of instability that are largely absent in conventional MRI where image quality depends on a stationary and homogeneous main field as well as stationary subject. Reorientation changes gravitational and inertial loading on the magnet and internal structures, while electrically conductive components in the magnet assembly can induce eddy currents during motion, together perturbing field homogeneity and temporal stability. Even when the subject and magnet are intended to move as a rigid assembly, neither living tissue nor the scanner components behave as perfectly rigid bodies. Relative motion between brain tissue and the encoding fields is therefore expected, even with firm head fixation, and will be particularly consequential for moving functional MRI, where the data are dependent on small signal differences, and are highly sensitive to misregistration and motion-induced phase errors.

The present mMRI system can support motion dynamics relevant to vestibular and sensorimotor paradigms, achieving net upward tilt rates up to 3.5°/s, while downward tilt rates up to 6.0°/s. For direct comparison, human roll tilt perception thresholds in complete darkness (i.e., without visual inputs) for most healthy people are less than 1°/s at 0.2 Hz[22]. Furthermore, yaw rotation thresholds at a frequency of 0.05 Hz have been reported to be 2.8°/s across a healthy human population[36], which is also within the motion range accessible to our platform. Perceptibility, however, depends on the full dynamic motion profile (amplitude, velocity, and acceleration): even slow periodic motion can be noticeable at sufficiently large amplitudes or acceleration. Static tilt is readily perceptible by humans, and likely for rats too given the similarity of the peripheral vestibular system in these two mammalian species. These considerations motivate careful selection of motion paradigms and tighter control of kinematic profiles in future implementations of mMRI.

Several practical observations from this prototype mMRI scanner inform both interpretation and next steps. First, the measured frequency shift versus tilt angle depended on the external magnetic environment: the $\Delta f(\theta)$ pattern differed between the 14 T ON and 14 T OFF conditions, indicating that external fields can contribute measurably to the observed frequency shift. Second, the relationship was slightly phantom-dependent, with water phantoms exhibiting less linear behavior than agar, consistent with increased motion-induced fluid dynamics. These findings suggest that both environmental fields and subject/phantom properties can modulate the effective frequency offset and may introduce departures from the simple linear model as motion speed increases. Frequency shifts in principle affect all three gradient axes, since slice position, frequency- and phase-encoding positions all depend on the frequency reference. Nevertheless, correcting projection displacement along the frequency-encoding direction alone substantially improved image quality, with readout-axis misregistration emerging as the dominant artifact mechanism under the present conditions.

A key enabling component of this implementation is the cryogen-free magnet architecture. By eliminating liquid cryogens and associated vessels, plumbing, and quench ducting, cryogen-free operation reduces risks associated with tilt and motion, and simplifies the mechanical design relative to conventional superconducting systems. Nevertheless, conduction-cooled magnets can exhibit cyclic field variations driven by cryocooler dynamics. While pulse-tube cryocoolers create less vibration than Gifford–McMahon cryocooler designs, their cooling performance can degrade under substantial tilt; the limited time at larger tilts in the present experiments mitigated this concern, but sustained or more aggressive motion will require explicit thermal and field-stability considerations.

The principal limitation of the current correction framework is its reliance on a spatially uniform, temporally stable $\Delta f(\theta)$ for each acquired k-space line. This assumption is violated by spatial gradients and higher-order field terms that arise during motion, as well as by any intra-line temporal variation in $\Delta\omega$, and these effects will become more prominent for faster motion and for pulse sequences with long readouts. In addition, slice-selection displacement cannot be corrected retrospectively, motivating either prospective frequency/slice tracking or acquisition strategies that are inherently less sensitive to slice-profile mismatch. More broadly, the present experiments do not yet address imaging modes most relevant to human neuroimaging during motion, such as echo-planar-imaging (EPI)-based fMRI, where sensitivity to phase errors, off-resonance, and rapid field changes is substantially higher.

Future work will therefore focus on quantifying the coupled mechanics–field dynamics of the platform across motion waveforms and speeds, and on extending correction from a global offset to spatially varying field models using field probes, navigators, or rapid $B_0$ mapping during motion. Prospective approaches that update transmit/receive frequency, shim terms, and spatial encoding in real time will likely be required for robust performance, particularly for EPI. Integrating real-time pose information about the subject (e.g., optical tracking or encoder feedback) with adaptive spatial encoding by the pulse sequence may further reduce blurring and ghosting by maintaining correct spatial encoding during motion, and would also enable systematic studies of susceptibility-driven field changes as head position evolves. Improvements in motion platform control, including tighter tolerance-based actuation and closed-loop controllers, will be important for reproducible kinematic profiles and for exploring motion regimes relevant to vestibular and sensorimotor paradigms.

A critical question to consider is if scaling mMRI to human use is at all possible. Achieving this will require engineering advances in magnet design, motion platform capability, and safety systems. The magnet assembly used here weighed 590 kg, whereas typical clinical MRI magnets weigh in the range of 5,000 – 10,000 kg. There are no purely conduction-cooled, completely liquid helium-free magnets on the market today. Even sealed magnets containing a small amount of liquid helium require a pressure vessel when the magnet is warm, and it remains unclear whether the pressure vessel could safely withstand large-scale motion. Environmental constraints also present challenges. Most clinical MRI systems are installed in RF-shielded, and often iron-shielded, rooms. Moving a magnet within an electrically conductive enclosure would induce substantial eddy currents, generating forces on both the surrounding structure and the magnet itself, as well as causing significant image distortion. In iron-shielded environments, motion could compromise shielding performance and, in extreme cases, lead to mechanical or structural failure.

Despite these challenges, no fundamental physical barriers preclude human-scale mMRI. The extremity system used here operates in an open laboratory without external RF or static magnetic shielding, supporting the feasibility of such configurations under less constrained conditions. Continued studies in small-animal models will be important for refining system design and understanding motion-related effects before translation to human-scale implementations.

## Materials and Methods

### *Extremity MRI Scanner*

A homebuilt extremity 1.5 T MRI scanner intended for imaging human arms and legs, small animals, and tissue samples was used. The magnet was made by Superconducting Systems Inc. (SSI, Billerica, MA, USA, acquired by IMRIS Company in 2022), and the scanner console was an RS2D (Reinvent Systems for Science and Discovery, Mundolsheim, France) Chameleon 3 multinuclear MR console. Conduction-cooled by an electrically powered cryocooler, the magnet eliminates the need for liquid helium or nitrogen (a.k.a. cryogen-free). The axial length of the bore is 550 mm and the center bore diameter is 300 mm, with 225 mm inside the gradient coil. All connections to the magnet, including electrical cables, cooling water and helium compressed gas hoses, are flexible and routed at the rear, allowing it to be tilted or moved while energized and with scanning in progress. The internal mechanical structure of the magnet is robust enough to allow movement and tilting while at field.

The same laboratory suite also housed an actively shielded 14 T (600 MHz) system (Bruker BioSpin Avance III; Magnex Scientific wide-bore vertical magnet). The extremity magnet and the 14 T magnet were located 5.91 m apart, as shown in **Fig. S2**. During the study period, the 14 T magnet was decommissioned, enabling data acquisition both while the magnet was energized ("14 T ON") and after it was de-energized ("14 T OFF").

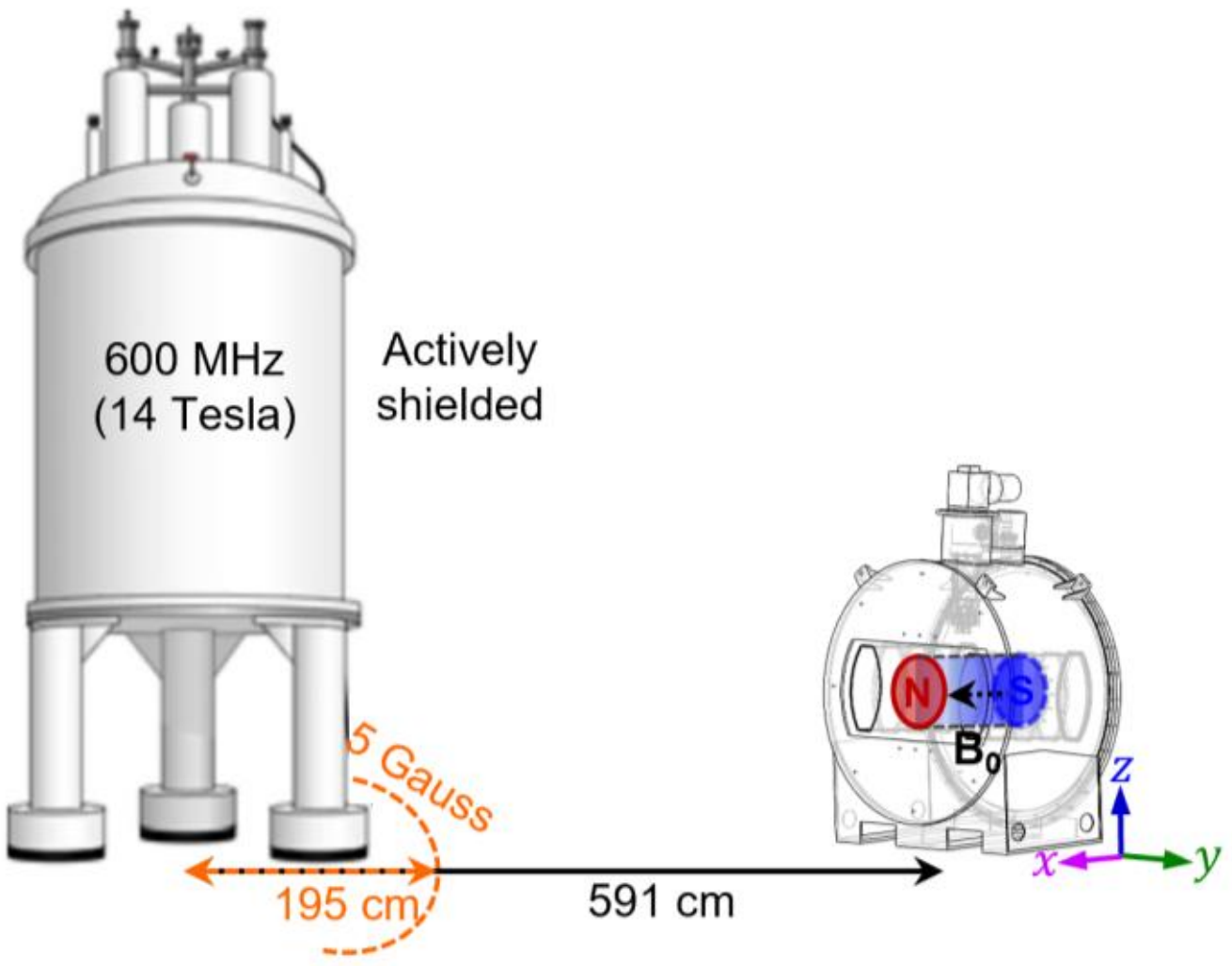

**Fig. S2 Site layout of the extremity magnet relative to the nearby 14 T magnet.** Distances are shown in a horizontal plane parallel to the floor. The center-to-center separation between the two magnets is 591 cm. The 14 T 5-gauss boundary is indicated, lying 195 cm from the magnet center along the line connecting the two magnets. $B_0$ denotes the main static field of the extremity magnet.

***Center of Gravity and Maximum Tilt Angle***

To minimize torques and moments on the magnet assembly, the extremity magnet should rotate as near as possible to its center of gravity (CoG). The CoG location also determines the maximum safe tilt angle in this configuration, which is defined by pivoting at the rear of the magnet stand, as implemented in this study. A multi-point weighing approach using a single strain gauge (FC23 compression load cell by TE Connectivity, Schaffhausen, Switzerland) was devised to measure the magnet weight and determine the CoG in three dimensions. In contrast to existing methodologies[37] that rely on rigid mounting platforms with multiple load cells, this method is cost-effective and straightforward to implement. Using this approach, the magnet weight was determined to be 590 kg. Details of the structural modelling and experimental measurements used to estimate the CoG and weight are presented in **Supplementary Materials A**.

### ***Motion Platform***

The pneumatically actuated tilt platform (**Fig. 1d-f**) was driven by a pair of round-body double-acting pneumatic cylinders with a 3-inch bore and a 10-inch stroke. These cylinders served as robust actuators, delivering a force of up to 1400 pounds-force at 100 psi using standard laboratory compressed air. To ensure controlled motion, airflow-limiting valves were used to separately regulate the angular rates of tilt during upward and downward movements. For safety and stability, nonmagnetic safety chains bolted to the floor constrained the tilt angle in the event of automatic control failure, improper manual operation, or structural failure of the pistons. Additional rubber cushioning was installed beneath the magnet to mitigate shock during rapid descent.

### ***Automatic Motion Control***

Magnet motion was controlled via an automated on/off feedback loop (**Fig. S3** and **Supplementary Movie 1** **link**). Key components included a LabJack T7 data acquisition system (LabJack Corporation, Lakewood, CO, USA) and a digital 3-axis accelerometer (ADXL345, Analog Devices, Wilmington, MA) to measure instantaneous tilt angle. The pulse sequence generated a transistor-transistor logic (TTL) trigger at each TR to synchronize image acquisition with LabJack data acquisition. The control loop actuated two pneumatic valves: one for pressurizing (intake) and one for exhausting the cylinders. When the measured tilt angle fell below the target, the intake valve was activated to extend the cylinders and increase the tilt angle. Conversely, when the tilt angle exceeded the target, the exhaust valve was activated to retract the cylinders against the magnet weight, reducing the tilt angle. After exceeding the target, the cylinders remained in position for one control cycle before retraction was initiated. Due to the angular momentum of the magnet assembly, the tilt angle often overshot the target. To mitigate this effect, a heuristic offset was applied by setting the control target a few degrees lower than the desired tilt angle. The control loop duration of 250 ms was determined based on the response time of the electrical valve (VEX3122-02N5DZ1-F 3-port 3-position center-off base-mounted proportional pneumatic valve, SMC Corporation, Tokyo, Japan). With this control scheme, tilt rates were regulated during both upward and downward motion.

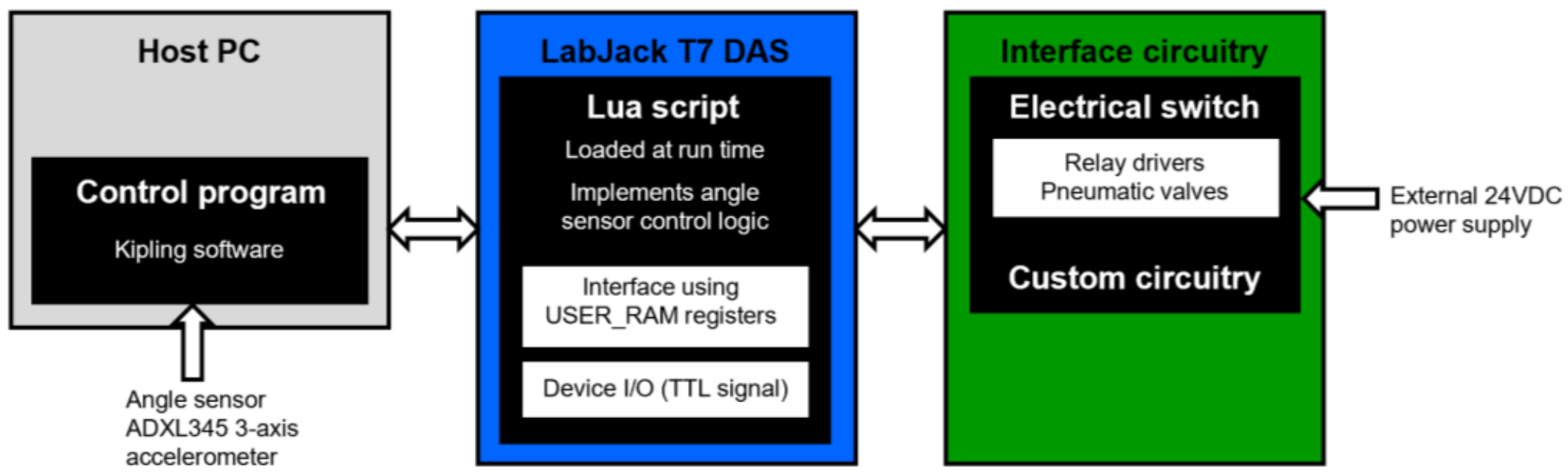

**Fig. S3 Schematic of the automatic motion control system.** Key components of the system included a host computer, LabJack data acquisition system (DAS), and electronic circuitry supporting the function of pneumatic valves. A custom Lua script was implemented in LabJack DAS for communication.

***Phantom mMRI Experiments***

A solenoidal RF transmit/receive coil was used to image a phantom of doped water (0.26% NaCl and 0.13% $NiSO_4$) filled in a 50 mL centrifuge tube (Corning, Inc., Corning, NY, USA; mean outer diameter 30 mm, mean inner diameter 28 mm). Seven 5 mm-diameter glass tubes were placed inside the centrifuge tube. The solenoidal coil was custom-made for proton imaging at 1.5 T, with an inductance of 350 nH and dimensions of 52 mm (length) and 37 mm (diameter). During scanning, the coil and phantom were securely positioned in the magnet. Single-slice spin-echo (SE) images were acquired with TR/echo time (TE) = 484/54 ms, matrix size = 256 x 256, field of view (FoV) = 50 x 50 mm$^2$, receiver bandwidth = 10 kHz, slice thickness = 5 mm. A single excitation was used per phase-encode line under both stationary and moving conditions.

***Anesthetized rat mMRI Experiments***

A single-loop, 28 mm diameter, surface coil resonating at proton frequency of 63.87 MHz was made for imaging the rat brain. The coil demonstrated a reasonable reflection coefficient, achieving -20 dB or lower. When loaded with a spherical phantom filled with doped water (0.26% NaCl and 0.13% $NiSO_4$) and of similar size to a rat's head, the coil exhibited a quality factor of 25. Before imaging, frequency tuning and impedance matching were performed on the RF coil using a Copper Mountain Technologies (Indianapolis, IN) TR1300/1 vector network analyzer. The magnetic field was shimmed to a linewidth of about 0.6 ppm. *In vivo* rat brain images were acquired using two-dimensional gradient-echo (GE) with TR/TE = 156/19 ms, matrix size = 128 x 128, FoV = 40 x 40 mm$^2$, slice thickness = 3 mm, 3 slices, slice spacing = 0.5 mm, bandwidth = 20 kHz and a single excitation for each phase-encode line.

Animal experiments were conducted in compliance with Institutional Animal Care and Use Committee (protocol #2022N000206) standards and the United States Department of Agriculture regulations. Healthy, Long-Evans rats (procured from Charles River Laboratories, Wilmington, Massachusetts, USA) were used for brain imaging *in vivo*. The rats were housed in the animal facility at Massachusetts General Hospital (Charlestown, Massachusetts, USA) and were given ample time (more than two weeks) to acclimate to the environment. At the time of the experiment, the rats were 15 - 18 weeks of age, weighed 350 - 410 grams with a head dimension of about 28 x 30 x 20 $mm^3$.

A tabletop veterinary anesthesia vaporizer (Smiths Medical SurgiVet, Minneapolis, MN, USA) was used to induce anesthesia with oxygen-enriched air (30% oxygen and 70% nitrogen; Airgas Inc., Radnor, PA, USA) containing 4% isoflurane, followed 1% isoflurane for maintenance at a flow rate of 7 cc/min supplied via a custom nose cone. To ensure proper hydration during the experiment, 0.9% sodium chloride (Baxter International Inc., Delaware PA, USA) was administered via injection. To secure the rat's head, a bite bar integrated into the nose cone was utilized. The temperature in the magnet was 22 °C, as measured by an HH64 (Omega Engineering, Norwalk, CT, USA) thermocouple readout. The rat was positioned on a heating pad containing flowing water at 37 °C from a pump (Stryker TP700, Kalamazoo, Michigan, USA). Prior to positioning the rat in the magnet, its respiratory rate was recorded at 60 breaths per minute. Respiration rate was monitored throughout the experiments. At the end of the study, the rat was euthanized, following American Veterinary Medical Association (AVMA) guidelines for the euthanasia of animals.

The rat and heating pad were supported in the magnet on a 3D printed cradle coated with epoxy resin for easy cleaning (**Fig. S4**). Atop this cradle, the surface RF coil's circuit board, with tuning and matching capacitors, was secured using plastic thumb screws. The RF coil was supported by the circuit board, centered above the rat's brain, and in contact with the head without exerting excessive pressure. The cradle was mounted on a custom plastic scissor jack, enabling precise vertical height adjustments to align the brain with the magnet isocenter. The jack assembly was mounted to two rings which locked within the magnet bore using cams to expand the rings slightly, affording the capability to establish precise static tilts around the rat's body axis.

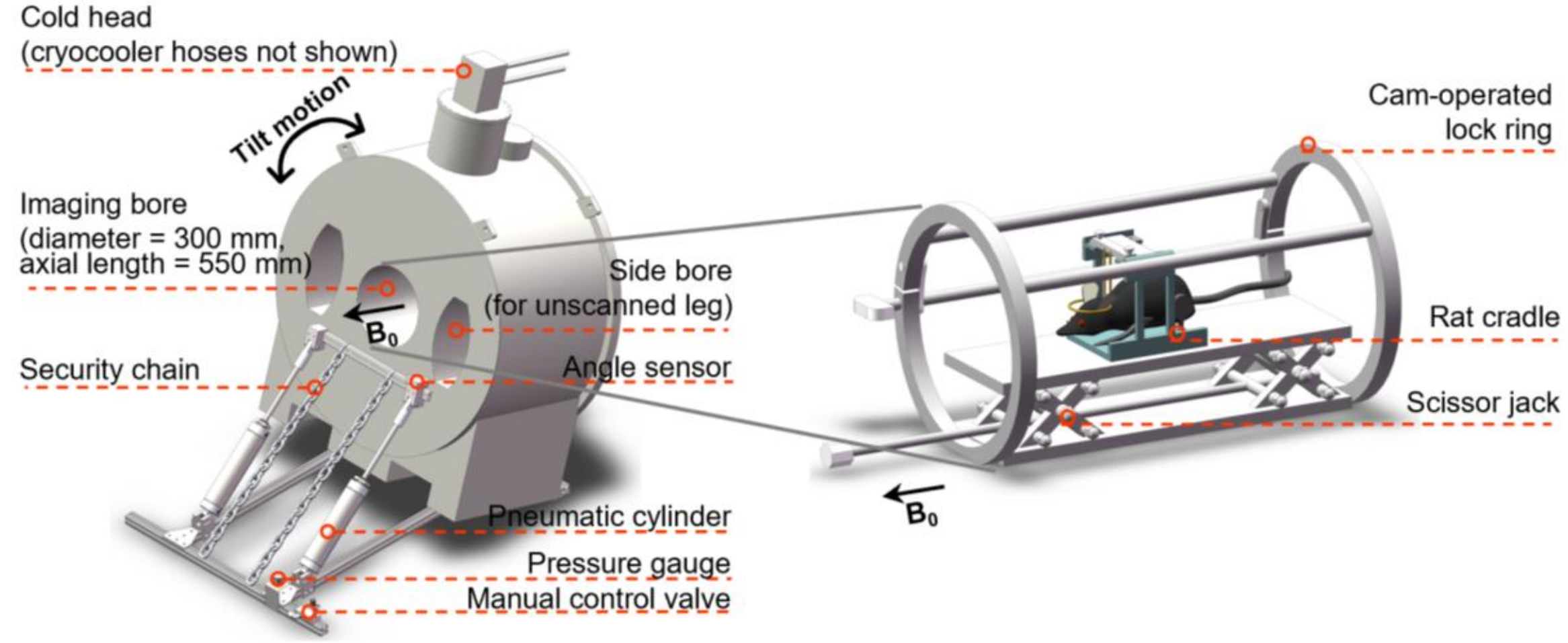

**Fig. S4 Main components for immobilizing the rat within the magnet bore.** The assembly includes the rat holder, scissor jack and locking rings. For clarity of presentation, details of the RF coil circuitry atop the rat cradle, as well as the nose cone and heating pad used in live animal experiments, are not shown.

***Correction for Tilt-Induced Frequency Shifts***

Mechanical tilting of the scanner introduces an angle-dependent resonance-frequency shift $\Delta\omega(\theta)$. In SE and GE scans of this study, each k-space line is acquired from a single echo at a fixed TE. For simplicity, we assume that a spatially uniform $\Delta\omega(\theta)$ influences only the frequency-encode dimension during the readout gradient ($G_x$), producing a deterministic phase evolution across the readout window. In image space, this linear readout-phase term is equivalent to a uniform translation of the reconstructed image along the frequency-encode axis, with displacement

$$\Delta x = \frac{\Delta\omega(\theta)}{\gamma G_x}, \quad (1)$$

where $\gamma$ is the proton gyromagnetic ratio.

The influence of $\Delta\omega(\theta)$ differs for SE and GE. In SE, the 180° pulse refocuses the pre-echo phase evolution, so the echo at TE remains phase-aligned in the rotating frame. However, $\Delta\omega(\theta)$ continues to accumulate phase during the readout interval and results in the shift $\Delta x$ along the frequency-encoding direction. In GE, $\Delta\omega(\theta)$ introduces both a phase offset at TE and the same readout-related displacement.

For retrospective correction, a single frequency shift was assigned to each acquired $k_y$ line. Specifically, for the $m$-th acquired k-space line associated with tilt angle $\theta(m)$, $\Delta\omega(\theta(m))$ (equivalently $\Delta f(k_y)$) was estimated using the static angle–frequency relationship reported in **Fig. 2c** (14 T ON) and **Fig. 2i** (14 T OFF), and was assumed to be effectively constant over the readout duration of that line. The readout samples were then retrospectively demodulated using the correction derived in **Supplementary Materials B**. For SE, the readout-induced phase term was removed by multiplying the measured signal by $e^{i\Delta\omega(\theta(m))t}$, where $t = 0$ was defined at the echo center. For GRE, an additional constant term $e^{i\Delta\omega(\theta(m))TE}$ was applied to remove the phase accumulated up to TE. The resulting k-space was then reconstructed using a conventional 2D Fourier transform.

Tilt-dependent frequency offsets also perturb the slice-selection. During excitation (and refocusing for SE), the RF pulse is applied concurrently with the slice-selection gradient $G_z$. A uniform $\Delta\omega(\theta)$ shifts the slice-selection profile along the slice-select axis by $\Delta z = \frac{\Delta\omega(\theta)}{\gamma G_z}$. Because this displacement occurs at the time of excitation and alters the spin population contributing to the signal, it cannot be corrected retrospectively by any k-space operation. When the tilt angle varies rapidly (or equivalently, for very long frequency-encoding readouts), the angle can change significantly during the readout, precluding retrospective correction. A first-order correction would require adding a quadratic term to the instantaneous phase correction factor based on the angular velocity estimated from adjacent phase-encode lines. Full derivations of SE and GE corrections are presented in **Supplementary Materials B**.

### *Mask-Optimized Artifact Correction*

Residual artifacts were further reduced by refining the nominal per-line frequency shifts, $\Delta f_{\text{nom}}(k_y)$, using a constrained greedy search strategy driven by an artifact mask. Masks $\Omega_{\text{art}}$ and $\Omega_{\text{sig}}$ (**Fig. 4d**) were held fixed across reconstructions for a given dataset. Candidate corrections were generated as $\Delta f(k_y) = \Delta f_{\text{nom}}(k_y) + \epsilon(k_y)$, where $\epsilon$ was sampled from a bounded discrete grid $\{-\Delta f_{max}, \ldots, -\Delta f_{step}, 0, \Delta f_{step}, \ldots, \Delta f_{max}\}$. When optimizing a given $k_y$ line, the candidate $\Delta f(k_y)$ was applied while holding all other lines fixed. The optimal frequency shift, $\Delta f^*(k_y)$, minimizes the Artifact-ROI intensity SD,

$$\sigma_{\text{art}} = \text{std}(|I|(\Omega_{\text{art}})), \quad (2)$$

where $|I|$ is the image magnitude. The Signal-ROI mean intensity,

$$\mu_{\text{sig}} = \text{mean}\big(|I|(\Omega_{\text{sig}})\big), \quad (3)$$

was computed as a secondary summary metric to monitor signal-level preservation.

This single-pass optimization is $k_y$ line order dependent, resembling coordinate-descent-like behavior, in which a single $k_y$ update alters the global image magnitude and $\sigma_{\mathrm{art}}$ is recomputed after each accepted update. Order dependency was examined by evaluating multiple update orders, including top-down traversal (monotonic in acquisition index), center-out traversal (starting from central $k_y$ and expanding outward), and random permutations. Multiple sweeps over all $k_y$ lines were performed so that earlier updates could be revisited after subsequent refinements. Random restarts, implemented as independent random traversals initialized from $\Delta f_{\mathrm{nom}}(k_y)$, were used to reduce sensitivity to local minima. All candidate frequency offsets were bounded to constrain the refinement to a local neighborhood around $\Delta f_{\mathrm{nom}}(k_y)$ and to prevent overcorrection. Optimization was terminated after a fixed number of sweeps or when further reductions in $\sigma_{\mathrm{art}}$ became negligible.

***Statistical Analysis***

Group-level analysis (**Fig. 7** and **Table 2**) used dataset-level averages, with slices from the same motion acquisition run averaged into a single observation for each dataset. Under 14 T ON conditions, N = 8 motion acquisitions were acquired with 3 slices each ($n_{\mathrm{slices}} = 24$) at TR = 50, 100, 150, 200, and 500 ms (N = 2, 1, 1, 1, and 3 acquisitions, respectively); under 14 T OFF conditions, N = 5 single-slice motion acquisitions were acquired at TR = 500 ms ($n_{\mathrm{slices}} = 5$). Pairwise comparisons between correction strategies were performed using a two-sided paired Wilcoxon signed-rank test, and Benjamini–Hochberg FDR correction was applied across the set of pairwise tests within each 14 T condition. A global Kruskal–Wallis test was used to assess whether $\sigma_{\mathrm{art}}$ differed among strategies within each condition (**Table 2**). Statistical significance was defined as $p < 0.05$.

## Acknowledgments

We thank Dr. Shahin Pourrahim, President of Superconducting Systems, Inc., (acquired by IMRIS in 2022), for providing detailed information on the extremity magnet. We are also grateful to Prof. Warren Warren for suggesting that external fields may be contributing to the observed artifacts.

**Funding:**

National Institutes of Health (NIH) National Institute of Mental Health (NIMH) and the National Institute of Biomedical Imaging and Bioengineering (NIBIB) grant R01EB029818 (JLA)

NIH NIBIB grant R01HD110152 (AVDK)

NIH National Institute of Child Health and Human Development grant R01HD099846 (ACDK)

NIH National Institute on Aging grant R01AA030014 (AVDK)

NIH National Institute of Neurological Disorders and Stroke grant U24NS135561 (AVDK)

The content is solely the responsibility of the authors and does not necessarily represent the official views of the NIH.

**Author contributions:**

Conceptualization: JLA, DMM

Methodology: JLA, DMM, JY, AK, AVDK, YCIC

Investigation: JLA, JY, AK, NP

Visualization: JY, AK

Supervision: JLA, DMM, AVDK, YCIC

Writing—original draft: JY, JLA

Writing—review & editing: DMM, AVDK, AK, YCIC

**Competing interests:** Authors declare that they have no competing interests.

**Data and materials availability:** All data supporting the findings of this study are available within the main text and supplementary materials. Additional data are available from the corresponding author upon reasonable request.

## Supplementary Materials

### A. Center of Gravity and Maximum Tilt Angle Calculation

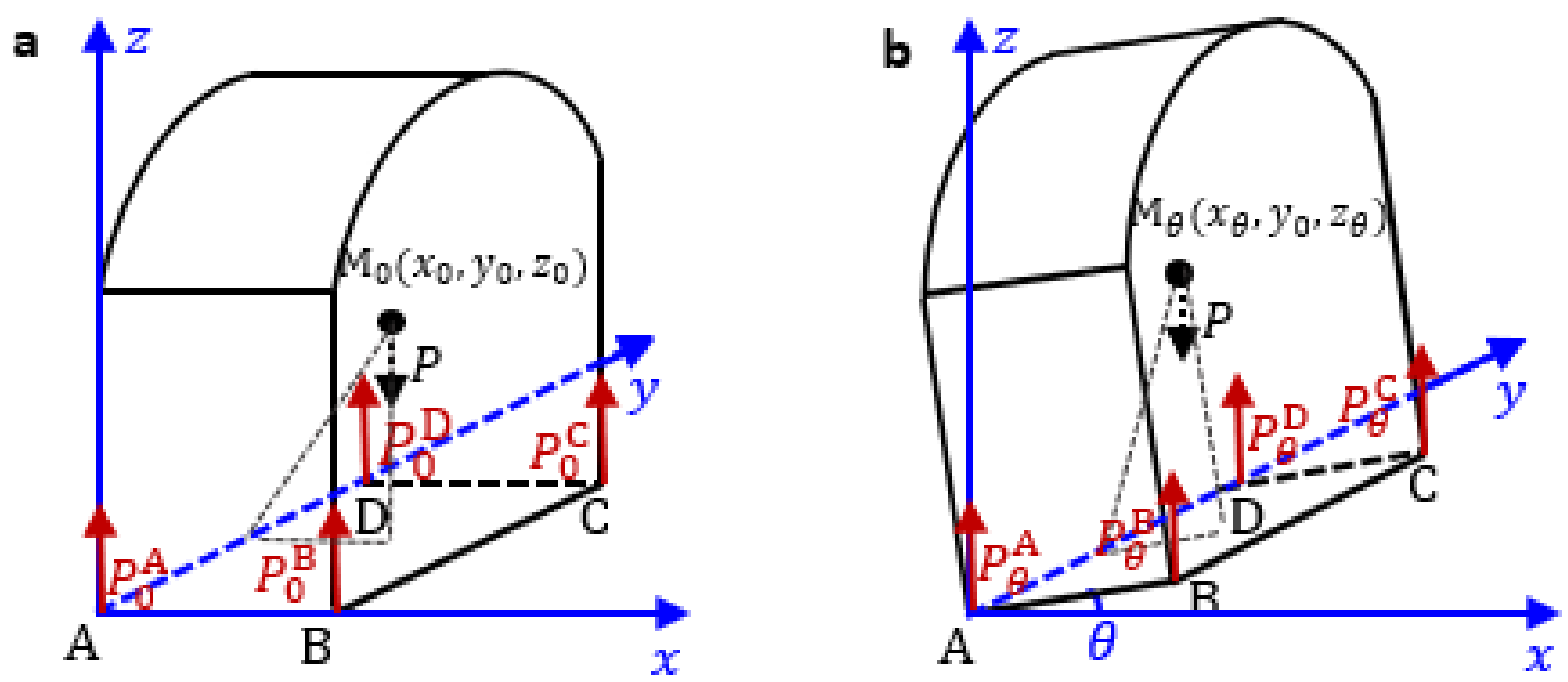

**Fig. S1 Free body diagram illustrating a simplified 3D representation of the magnet's structure. a** Magnet in its untilted position. **b** Magnet tilted by an angle $\theta$ around edge $\mathrm{AD}$.

The procedures used to determine the weight, CoG, and maximum tilt angle of the magnet are outlined. For mathematical modelling purposes, the MRI apparatus was simplified to a three-dimensional structure (**Fig. S1**), with its projection on the ground forming a parallelogram denoted by the vertices $\mathrm{A}$, $\mathrm{B}$, $\mathrm{C}$, and $\mathrm{D}$, and corresponding support forces $P^{\mathrm{A}}$, $P^{\mathrm{B}}$, $P^{\mathrm{C}}$, and $P^{\mathrm{D}}$ at these points. When the magnet is not tilted, the support forces are $P_0^{\mathrm{A}}$, $P_0^{\mathrm{B}}$, $P_0^{\mathrm{C}}$ and $P_0^{\mathrm{D}}$, and position of the CoG is denoted as $\mathrm{M}_0(x_0, y_0, z_0)$. The following relationship is established based on the first law of equilibrium:

$$\begin{cases} (P_0^{\mathrm{A}} + P_0^{\mathrm{D}})x_0 = (P_0^{\mathrm{B}} + P_0^{\mathrm{C}})(l_{\mathrm{AB}} - x_0) \\ (P_0^{\mathrm{A}} + P_0^{\mathrm{B}})y_0 = (P_0^{\mathrm{C}} + P_0^{\mathrm{D}})(l_{\mathrm{AD}} - y_0) \end{cases} \quad (1)$$

where $l_{\mathrm{AB}} = 54.7$ cm is the length of edge $\mathrm{AB}$, $l_{AB} = 101.5$ cm is the length of edge $\mathrm{AD}$, and $P = P_0^{\mathrm{A}} + P_0^{\mathrm{B}} + P_0^{\mathrm{C}} + P_0^{\mathrm{D}}$ represents the weight of the magnet. Placing the load cell at each vertex, we quantified individual support forces. Based on **Eqs. (1)**, the calculated CoG within the $(x_0, y_0)$ plane is $(21.7, 50.3)$ cm, and the weight of the magnet is 590 kg.

When the magnet is tilted by a known angle $\theta$ around edge $\mathrm{AD}$ (**Fig. S1b**), the support forces at the four points shift to new values denoted by $P_\theta^{\mathrm{A}}$, $P_\theta^{\mathrm{B}}$, $P_\theta^{\mathrm{C}}$, and $P_\theta^{\mathrm{D}}$. The CoG also shifts to a new position denoted by $\mathrm{M}_\theta(x_\theta, y_0, z_\theta)$. Again, based on the first law of equilibrium, we derive:

$$(P_\theta^{\mathrm{A}} + P_\theta^{D})x_\theta = (P_\theta^{\mathrm{B}} + P_\theta^{\mathrm{C}})(l_{\mathrm{AB}}\cos\theta - x_\theta) \quad (2)$$

Similarly, using the load cell, $P_\theta^{\mathrm{A}}$, $P_\theta^{\mathrm{B}}$, $P_\theta^{\mathrm{C}}$, and $P_\theta^{\mathrm{D}}$ were measured, allowing for calculating $x_\theta$. In our experiment, we tilted the magnet by $\theta = 8.36^\circ$ which yielded $x_\theta = 11.1$ cm. Furthermore, the following geometric relationships hold:

$$\begin{cases} x_0^2 + z_0^2 = x_\theta^2 + z_\theta^2 \\ \cos^{-1}\left(x_0/\sqrt{x_0^2 + z_0^2}\right) + \theta = \cos^{-1}\left(x_\theta/\sqrt{x_\theta^2 + z_\theta^2}\right) \end{cases} \tag{3}$$

With calculated $x_0$ and $x_\theta$, the unknowns in **Eqs. (3)** are $z_0$ and $z_\theta$. Introducing $\varphi = \cos^{-1}(x_0/\sqrt{x_0^2 + z_0^2})$ and $\delta^2 = x_0^2 + z_0^2 = x_\theta^2 + z_\theta^2$ provides $\cos\varphi = x_0/\delta$ and $\cos(\varphi + \theta) = x_\theta/\delta$. This simplifies to $\frac{\cos\varphi}{x_0} = \frac{\cos(\varphi+\theta)}{x_\theta}$. Let $f(\varphi) = x_\theta \cos\varphi - x_0 \cos(\varphi + \theta)$. Solving for $f(\varphi) = 0$ provides $z_0$. While the analytical solution for $z_0$ is intractable, a practical estimate of $z_0$ was obtained by identifying the zero-crossing point $\varphi_0$, yielding $z_0 = 69.9$ cm.

In summary, the CoG in the established coordinate system is $(21.7, 50.3, 69.9)$ cm and the critical angle for tilting without toppling over is $\theta_c = \tan^{-1}(x_0/z_0) \approx 17.28^\circ$.

## B. Retrospective Correction for Tilt-Dependent Frequency Offsets

This appendix provides the mathematical formulation underlying the correction applied to spin-echo (SE) and gradient-echo (GE) data in the presence of a spatially uniform, tilt-dependent resonance-frequency shift $\Delta\omega(\theta)$. Throughout, $\theta(m)$ denotes the scanner tilt angle at the time the $m$-th k-space line is acquired. Throughout the main text, frequency shifts are reported as $\Delta f = \Delta\omega/2\pi$ ; this appendix uses angular frequency $\Delta\omega$ for compactness, and $\gamma$ denotes the gyromagnetic ratio in rad/s/T. In the implementation used here, $\Delta\omega(\theta)$ is treated as constant over the readout of each acquired $k_y$ line; i.e., $\Delta\omega(\theta(m))$ is sampled once per line and applied uniformly across its readout samples.

### *B.1 Signal Model During Readout*

During the frequency-encoding readout gradient $G_x$, the precession frequency at position $x$ is

$$\omega(x;\theta) = \gamma G_x x + \Delta\omega(\theta). \tag{4}$$

The acquired signal for the m-th k-space line is therefore

$$s_m(t) = \int \rho(x) \exp[-i(\gamma G_x x + \Delta\omega(\theta(m)))t] dx, \tag{5}$$

with $t = 0$ defined at the echo center. Introducing k-space coordinate along the frequency-encoding direction $k_x(t) = \gamma \int_0^t G_x(\tau) d\tau$ yields

$$s_m(k_x) = \exp[-i \frac{\Delta\omega(\theta(m))}{\gamma G_x} k_x] S_0(k_x, k_y^{(m)}), \tag{6}$$

where $S_0(k_x, k_y^{(m)})$ represents the untilted, on-resonance k-space signal.

### *B.2 Frequency-Encoding Distortion*

The multiplicative term $\exp[-i\Delta\omega(\theta)t]$ introduces a linear phase ramp across the readout, which appears in the reconstructed image as a spatial displacement

$$\Delta x = \frac{\Delta\omega(\theta)}{\gamma G_x}. \tag{7}$$

Both SE and GE acquisitions are subject to the readout-induced displacement in **Eq. (7)**.

### *B.3 Phase at TE in SE and GRE*

In SE, phase accumulated before the 180° pulse is inverted and refocused, so a spatially uniform $\Delta\omega(\theta)$ produces no residual phase at TE, and **Eq. (7)** is therefore the correctable in-plane displacement. In GE, the $\Delta\omega$-dependent phase that accumulates up to TE is not refocused and contributes an additional echo-phase term with magnitude:

$$\phi_{GRE}(\theta) = \Delta\omega(\theta)TE, \quad (8)$$

appearing in the signal as $\exp[-i\phi_{GRE}(\theta)]$, to the linear phase evolution $\exp[-i\Delta\omega(\theta)t]$.

### *B.4 Slice-Selection Displacement*

During excitation (and during the refocusing pulse for SE), the RF pulse is applied under a slice-selection gradient $G_z$. The effective resonance condition is shifted by $\Delta\omega(\theta)$, displacing the excited slice along the slice-select axis by

$$\Delta z = \frac{\Delta\omega(\theta)}{\gamma G_z}. \quad (9)$$

This displacement modifies the physical location of the spins that contribute to the echo. Because the error occurs at excitation and alters the prepared magnetization itself, it cannot be compensated retrospectively by any k-space operation. Slice-selection offsets therefore constitute an inherent source of spatial mismatch in the presence of $\Delta\omega(\theta)$.

### *B.5 Retrospective Correction*

For SE, removing the readout-induced phase term yields

$$\tilde{s}_m(k_x) = s_m(k_x)\exp[+i\Delta\omega(\theta(m))t]. \quad (10)$$

For GE, both the TE-dependent phase and the readout-dependent phase must be removed:

$$\tilde{s}_m(k_x) = s_m(k_x)\exp[+i\Delta\omega(\theta(m))TE]\exp[+i\Delta\omega(\theta(m))t]. \quad (11)$$

Substitution of **Eq. (6)** shows that $\tilde{s}_m(k_x) = S_0(k_x, k_y^{(m)})$. In discrete form, the correction can be applied per $k_y$ line as a multiplicative vector across readout samples, combining the TE phase factor (GE only) and the readout demodulation term implied by **Eq. (6) – (7)**.